\documentclass[twocolumn,secnumarabic,amssymb,superscriptaddress,nobibnotes, aps, prd]{revtex4-2}

\usepackage{comment}
\usepackage{graphicx}
\usepackage[charter,greekuppercase=italicized]{mathdesign}
\usepackage{amsmath, amsfonts, amssymb}
\usepackage{hyperref}
\usepackage{xcolor}
\usepackage{float}
\usepackage{placeins}
\usepackage[export]{adjustbox}
\hypersetup{
    colorlinks = true,
    linkcolor = blue,
    anchorcolor = blue,
    citecolor = blue,
    filecolor = blue,
    urlcolor = blue
    }   
\usepackage{multirow}
\usepackage{hhline}

\begin{document}

%%%%%%%%%%%%%%%%%%%%%%%%%%% Title %%%%%%%%%%%%%%%%%%%%%%%%%%%%%%%%%%%%
\title{Disorder-driven coexistence of distinct dynamical states in frustrated Sr$_3$CuNb$_2$O$_9$: a microscopic $\mu$SR and $^{93}$Nb NMR study}

\author{M. Biswas}
\address{Department of Physics, Shiv Nadar Institution of Eminence, Gautam Buddha Nagar, UP 201314, India}

\author{K. Bhattacharya}
\address{Department of Physics, Shiv Nadar Institution of Eminence, Gautam Buddha Nagar, UP 201314, India}

\author{K. M. Ranjith}
\address{Leibniz Institute for Solid State and Materials Research Dresden, 01069 Dresden, Germany}
\address{Würzburg-Dresden Cluster of Excellence ct.qmat, Technische Universität Dresden, 01069 Dresden, Germany}

\author{S. M. Hossain}
\address{Department of Physics, Shiv Nadar Institution of Eminence, Gautam Buddha Nagar, UP 201314, India}

\author{S. S. Islam} 
\address{Laboratory for Muon-Spin Spectroscopy, Paul Scherrer Institut, CH-5232 Villigen PSI, Switzerland}

\author{M. Naskar}
\address{Institute for Solid State and Materials Physics, Technische Universit¨at Dresden, D-01069 Dresden, Germany}

\author{R. Sarkar}
\address{Institute for Solid State and Materials Physics, Technische Universit¨at Dresden, D-01069 Dresden, Germany}

\author{B. Büchner}
\address{Leibniz Institute for Solid State and Materials Research Dresden, 01069 Dresden, Germany}
\address{Würzburg-Dresden Cluster of Excellence ct.qmat, Technische Universität Dresden, 01069 Dresden, Germany}

\author{T. Shiroka}
\address{Laboratory for Muon-Spin Spectroscopy, Paul Scherrer Institut, CH-5232 Villigen PSI, Switzerland}

\author{H.-J. Grafe}
\address{Leibniz Institute for Solid State and Materials Research Dresden, 01069 Dresden, Germany}

\author{M. Majumder}
\email[Contact author: ]{mayukh.majumder@snu.edu.in}
\address{Department of Physics, Shiv Nadar Institution of Eminence, Gautam Buddha Nagar, UP 201314, India}

\date{\today}

%%%%%%%%%%%%%%%%%%%%%%%%%%%%%%%%% Abstract %%%%%%%%%%%%%%%%%%%%%%%%%%%%%%%%
\begin{abstract}
Despite recent progress in identifying the exotic random singlet (RS) state in disordered frustrated magnets as a distinct correlated phase, three-dimensional (3D) realizations remain scarce. Sr$_3$CuNb$_2$O$_9$ was proposed to be one of such 3D frustrated systems with magnetic site disorder hosting an RS ground state. Here, we report a detailed microscopic investigation of  Sr$_3$CuNb$_2$O$_9$ employing muon spin relaxation ($\mu$SR) and $^{93}$Nb nuclear magnetic resonance (NMR) techniques. The $\mu$SR zero-field relaxation rate reveals a power-law divergence of the relaxation rate as a function of temperature. Also, a power-law divergence is present in the relaxation rate as a function of applied longitudinal field, consistent with the formation of an RS phase. The $^{93}$Nb NMR spectra unambiguously resolve two components with distinct local magnetic environments, whose nature is further elucidated through spin-lattice relaxation measurements analyzed via an inverse Laplace transform (ILT) of the nuclear magnetization recovery. The relaxation-rate distribution obtained from ILT reveals two well-separated channels: a fast component, $(1/T_1)_{\mathrm{fast}}$, and a slow component, $(1/T_1)_{\mathrm{slow}}$. Both components follow distinct power-law temperature dependences ($T^{\alpha}$), with $\alpha = 0.6$ and $1.1$ for the fast and slow channels, respectively. The combined spectral and relaxation data demonstrate that the fast channel qualitatively represents an RS-like state, whereas the slow channel exhibits quantum spin liquid (QSL) like behavior, thereby establishing the microscopic coexistence of RS and QSL-like phases in Sr$_3$CuNb$_2$O$_9$.

\end{abstract}

\maketitle

%%%%%%%%%%%%%%%%%%%%%%%%%%%%%%%% INTRODUCTION %%%%%%%%%%%%%%%%%%%%%%%%%%%%%%
\section{INTRODUCTION}

Quenched randomness was long regarded as a weak perturbation that merely suppresses long-range magnetic order or quantum entanglement, with only a marginal impact on universal behavior. However, over the past two decades, it has become clear that disorder can drive intrinsically non-perturbative physics, fundamentally giving rise to phases distinct from clean critical points \cite{PhysRevLett.95.206603,PhysRev.109.1492}.  Among them, the random singlet (RS) state stands out for its distinctive nature: an extended network of singlet pairs formed over a wide distribution of length and energy scales, originating from an underlying power-law distribution of exchange couplings as first shown by Dasgupta and Ma   through strong-bond decimation leading to an infinite randomness fixed point (IRFP) \cite{PhysRevB.22.1305}. Later on, based on the IRFP framework, Bhatt and Lee predicted RS behaviors in higher dimensions \cite{PhysRevLett.48.344,10.1063/1.329684}. Furthermore, Kimchi \textit{et al.} recently advanced a theoretical framework in which unconventional low-temperature scaling behaviors in magnetization and heat capacity observed in certain disordered frustrated quantum spin liquid (QSL) candidates are interpreted as signatures of the RS phase~\cite{Kimchi2018}. Interestingly, T. Imai \textit{et al.} have demonstrated through the nuclear quadrupole resonance (NQR) measurements, a broad distribution of local spin environments and low-temperature components consistent with the formation of random spin singlets coexisting with residual paramagnetic spins, rather than a purely homogeneous QSL state in Zn-barlowite and herbertsmithiten - systems with disordered magnetic sites that were initially regarded as QSL candidates \cite{Wang2021}. Such phase coexistence has also been microscopically observed in several other systems \cite{PhysRevB.110.064418, PhysRevB.107.014424, chen2025spinonssolitonsrandomsinglets}.

Interestingly, most of these compounds are two-dimensional in nature, and comparatively little attention has been devoted to three-dimensional realizations of RS physics. Notably, a recent study by few of us established a spin-$\tfrac{1}{2}$ quasi-cubic frustrated system Sr$_3$CuNb$_2$O$_9$ (SCNO) as a compelling three-dimensional RS candidate, where both $ C/T $ and $ \chi $ exhibit power-law behavior $ T^{-\gamma} $ with $ \gamma \approx 0.6 \pm 0.05 $ and display striking data-collapse properties~\cite{PhysRevB.110.L020406}. Numerical modeling further supports a broad, inhomogeneous exchange distribution $ P(J) \sim J^{-\gamma} $, naturally arising from Cu/Nb site sharing and associated bond randomness. In this context, utilizing microscopic tools is compelling to understand the underlying mechanism of the randomness and the frustration to stabilize the ground state and possible phase coexistence in SCNO. 

In this work, we report a microscopic investigation of the magnetic ground state of SCNO using two complementary techniques, $\mu$SR and NMR. The zero-field $\mu$SR measurements reveal an RS ground state down to $ T = 0.4~\mathrm{K} $ with no evidence of long-range magnetic ordering or spin glass behavior, while longitudinal-field $\mu$SR further confirms the predominantly dynamic character of this state. The $^{93}\mathrm{Nb}$ NMR spectra and spin-lattice relaxation rate ($ 1/T_1 $) uncover clear signatures of phase coexistence, indicating the simultaneous presence of RS and QSL-like components at low temperatures down to 1.8~K. Taken together, these results provide direct microscopic insight into a quantum-disordered ground state stabilized by the interplay of frustration and randomness in this three-dimensional quantum magnet.

%%%%%%%%%%%%%%%%%%%%%%%%%%%%%%%%%%%%%%%%%%%%%%%%%%%%%%%%%%%%%%%%%%%%%%%%%%%%%%%%%%%%%%%%%%%%%%%%%%%%%%%%%%%%%%%%%%%%%%%%%%%%%%%%%%%%%%%%%%%%%%%%%%%%%%%%%%%

\section{EXPERIMENTAL DETAILS}
In this study, we perform experiments on polycrystalline compound SCNO \cite{PhysRevB.110.L020406}. We have utilized the $\mu$SR and NMR techniques to probe the static and dynamic properties of Sr$_3$CuNb$_2$O$_9$. In NMR, the nuclei probe the local electronic magnetic moments through hyperfine interactions, providing information on the static and dynamic spin susceptibility at the nuclear sites. In contrast, $\mu$SR involves anisotropy of emitted positrons from implanted spin-polarized muons situated at the interstitial position of a material, where they interact with the local magnetic environment. The zero-field (ZF) $\mu$SR measurement in the temperature range $100 - 0.4~$K and longitudinal field (LF) measurement with field values $\mu_0H = 0.015 - 0.5~$T have been measured using the Dolly spectrometer placed in the $\pi$E1 area of the Paul Scherrer Institute (PSI), Switzerland.

$^{93}\text{Nb}$ ($I=9/2$, $\gamma_{N}/2\pi= 10.407$ MHz/T) NMR measurements were performed using a phase-coherent Tecmag Redstone spectrometer equipped with a $16 \text{ T}$ superconducting magnet and a variable-temperature insert (VTI). NMR spectra were obtained as a function of frequency at various temperatures using a standard spin-echo pulse sequence ($\pi/2 - \tau - \pi$). For the frequency sweep NMR spectra, the resonant circuit tuning and impedance matching were achieved via computer-controlled stepper motors. The intrinsic NMR spectra were then reconstructed by summing the Fourier transforms of the collected spin-echo signals. The nuclear spin-lattice relaxation rate, $1/T_1$, was measured using a saturation-recovery pulse sequence with negligible heating effects. The time evolution of the nuclear magnetization, $M(t)$, was monitored by varying the recovery delay time ($t$) between the initial saturation (or $\pi/2$) pulse and the subsequent $\pi/2 - \tau - \pi$ spin-echo sequence. $M(t)$ was directly proportional to the integrated magnitude of the spin echo. Crucially, to ensure accurate measurement and  to capture the fast-relaxing $^{93}\text{Nb}$ signal, a very short time $\tau$ (the $\pi/2$ to $\pi$ pulse separation) of $3~\mu s$ and $5~\mu s$ was employed consistently for spectral and relaxation measurements, respectively.

%%%%%%%%%%%%%%%%%%%%%%%%%%%%%%%%%%%%%%%%%%%%%%%%%%%%%%%%%%%%%%%%%%
\begin{figure*}
\includegraphics[scale=0.35]{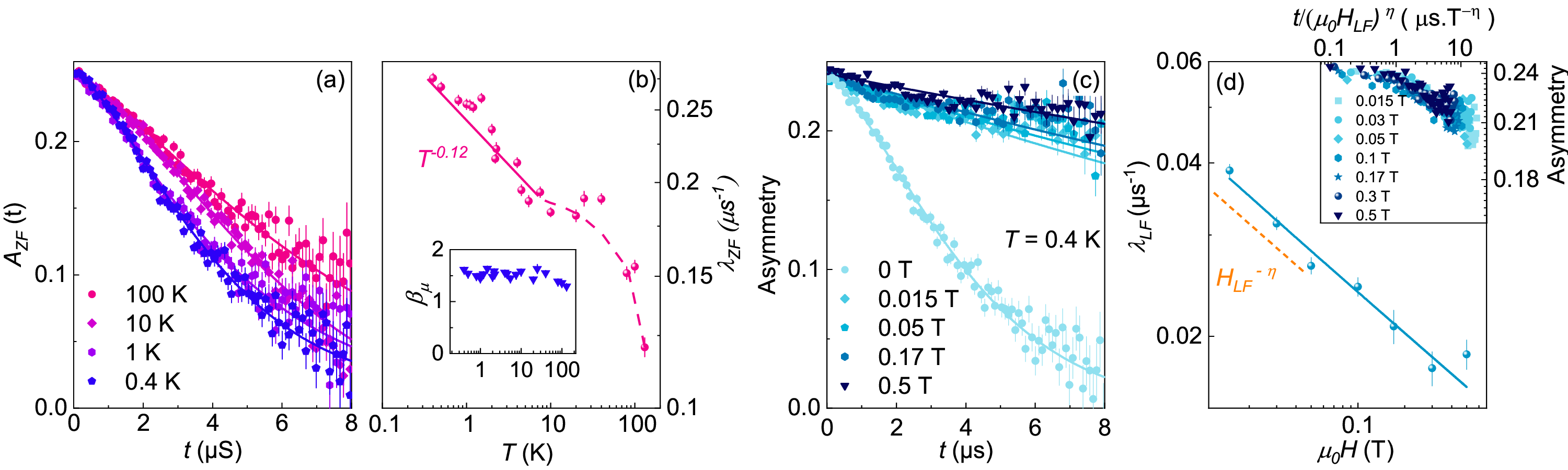}
\caption{\label{fig1}(a) ZF-$\mu$SR asymmetry spectra as a function of time for different temperatures. Solid lines are the fits using Eq.~\eqref{musr_eqn_ZF}. (b) ZF relaxation rate ($\lambda_{\rm ZF}$) vs $T$ exhibiting a power law $T^{- \gamma}$, where $\gamma=0.12$ and the dashed line represents the guide to the eyes. In the inset of (b) $\beta_\mu$ as a function of $T$. (c) $\mu$SR asymmetry spectra measured in different longitudinal fields at $T = 0.4$~K. (d) The corresponding relaxation rate $\lambda_{\rm LF}$ as a function of the longitudinal field and the solid line represents the fits using Eq.~\eqref{musr_Lambda_LF_eqn}. The dashed line represents  a power law $\lambda_{LF} \sim (\mu_0 H)^ \eta$ with $\eta = 0.26$. Inset of (d) depicts time-field scaling of longitudinal field muon asymmetry at $T =0.4~$K.}
\end{figure*}
%%%%%%%%%%%%%%%%%%%%%%%%%%%%%%%%%%%%%%%%%%%%%%%%%%%%%%%%%%%%%%%%%%

\section{RESULTS AND DISCUSSIONS}
\subsection{$\mu$SR}
The zero-field asymmetry measurement of $\mu$SR is one of the best approaches to detect any long-range ordering or any spin freezing. Fig.~\ref{fig1}(a) depicts the ZF asymmetries as a function of time at different temperatures. The system neither possesses long-range magnetic ordering nor static magnetic moments down to $T=0.4~K$ as these asymmetries do not exhibit any oscillations or $1/3$ polarization recovery in the greater time scale, respectively \cite{PhysRevB.31.546, PhysRevLett.73.3306}. To analyze these time-resolved asymmetry spectra, we modeled them with a stretched exponential function, which phenomenologically describes the behavior of a disordered system \cite{PhysRevMaterials.5.014411,Mustonen2018,PhysRevLett.74.3471, PhysRevB.96.014432},

\begin{equation}
\label{musr_eqn_ZF}
A_{ZF}(t) = A_0 ~ exp(-\lambda_{ZF} t)^{\beta_\mu}
\end{equation}

Here $ A_{0} $ is the initial muon asymmetry, $ \lambda_{\mathrm{ZF}} $ the zero-field muon relaxation rate, and $ \beta_{\mu} $ the stretching exponent. The ZF asymmetry exhibits a Gaussian-like shoulder only at early times ($ t < 2~\mu\text{s} $), and the fits yield $ 1 < \beta_{\mu} < 2 $, as shown in the inset of Fig.~\ref{fig1}(b). This behavior points to a distribution of relaxation channels corresponding to an effective admixture of Gaussian ($ \beta_{\mu} = 2 $) and Lorentzian ($ \beta_{\mu} \approx 1 $) local-field distributions. Where the former represents a dense disordered system and the latter represents a diluted disordered system \cite{Muon_Book_1}. The rationale for selecting a more homogeneous Gaussian distribution by the system is that the unpaired spins sporadically migrate from site to site within the time frame of the muon lifetime through the alternation of singlet bonds, resulting in greater homogeneity than a Lorentzian system \cite{PhysRevLett.73.3306, PhysRevMaterials.5.014411, PhysRevLett.127.157204}.  At high temperature $T > 100$~K, one can calculate the spin fluctuation frequency, $\nu = \sqrt{z}J_{ex} s/ \hbar \sim 6.41 \times 10^{12}$~Hz \cite{PhysRevB.110.L060403}, where $z=6$ is the nearest-neighbor coordination number for a cubic system and $J_{ex}/k_B = - 3 \theta_{CW}/zs(s+1) \sim 40$~K (with spin value $s=1/2$, and Curie-Weiss temperature $\theta_{CW} \simeq -60$~K \cite{PhysRevB.110.L020406}). Such a high spin fluctuating frequency reflects the uncorrelated paramagnetic fluctuations at high temperatures. With decreasing temperature from the high-temperature regime, the zero-field relaxation rate $ \lambda_{\mathrm{ZF}} $ increases, signaling the progressive development of short-range spin correlations. Below approximately 15~K, $ \lambda_{\mathrm{ZF}} $ follows a power-law behavior $ \lambda_{\mathrm{ZF}} \sim T^{-\gamma} $ with $ \gamma \approx 0.12 $ for $ T < 10~\mathrm{K} $, in sharp contrast to the nearly temperature-independent low-temperature relaxation typically observed in quantum spin liquid (QSL) candidates, where $ \lambda_{\mathrm{ZF}} $ tends to saturate at the lowest temperatures \cite{PhysRevB.110.L060403, PhysRevLett.125.267202, PhysRevLett.117.097201}. Noteworthy is that such divergence is generally observed in systems where the random distribution of exchanges is present \cite{PhysRevLett.126.037201, PhysRevB.110.064418, Khatua2022, PhysRevMaterials.5.014411, rw5c-491w}. This presence of the RS state in SCNO corroborates with the bulk measurements also \cite{PhysRevB.110.L020406}. 

It should be pointed out that the ZF-relaxation rate $\lambda_{ZF}$ increases to $\sim 0.28~ \mu S^{-1}$ (at $T \leq 10~K$) with a few values from $\sim 0.12~ \mu S^{-1}$ (at $T \geq 10~K$) by cooling down the system, on the contrary of spin-glass type transition whereas the relaxation rate increases with a few orders of magnitude \cite{PhysRevB.31.546, PhysRevLett.73.3306}. Also, in contrast to the spin glass scenario where the predicted value of $\beta$ below spin freezing temperature is 1/3 \cite{PhysRevB.32.7384, PhysRevLett.72.1291}, here the $\beta$ value is almost constant at $\sim 1.5 $ throughout all the temperatures as shown in the inset of Fig.\ref{fig1}(b).

Muon decoupling experiment using a longitudinal applied magnetic field in the direction of the initial muon direction is the most effective technique to determine the nature of the spin correlations. Fig.~\ref{fig1}(c) shows all the LF asymmetries as a function of time at different applied longitudinal fields measured at $T=0.4$~K. All the asymmetries fitted using an exponential function $A_{LF}(t)=A_0~ exp(-\lambda_{LF} t)$, where $A_0$ is the initial muon asymmetry and $\lambda_{LF}$ is the LF relaxation rate. Considering the relaxation is governed by static correlations, one can estimate the local static field $\langle B_{loc} \rangle \sim 0.33$~mT (where $\langle B_{loc} \rangle =\lambda_{ZF,~T=0.4~K} / \gamma_\mu$, with $\gamma_\mu/2\pi = 135.5~ s^{-1} \mu T^{-1}$). Usually, an LF five times the $\langle B_{loc} \rangle$ is sufficient to completely decouple the relaxation channel; however, the decoupling has not been observed up to an LF of 0.5~T, indicating the presence of dynamical correlations.

%The $\lambda_{LF}$ has a finite value even after applying 0.5~T of longitudinal field. If the relaxation process is associated with static magnetic moments (e.g. nuclear or electronic originating) then 33~mT of the longitudinal field would have been enough to fully decouple the muons, which is 100 times more than the local field develops by the static moments $\langle B_{loc} \rangle \sim 0.33$~mT (where $\langle B_{loc} \rangle =\lambda_{ZF,~T=0.4~K} / \gamma_\mu$, with $\gamma_\mu/2\pi = 135.5~ s^{-1} \mu T^{-1}$). 

Fig.~\ref{fig1}(d) shows $\lambda_{LF}$ as a function of applied longitudinal magnetic fields, which was inadequate to fit using the well-known Redfield fit \cite{PhysRevB.20.850} as in low-temperature the auto-correlation function $S(t) \sim (\tau / t)^x exp(- \nu t)$ (where x is the critical exponent, $\tau$ and $1/\nu$ are the early and late time cutoffs respectively) develops differently with $x\neq0$ rather a simple exponential function with $x=0$. To describe $\lambda_{LF}(\mu_0H)$ in an efficient way we have used the formula,

\begin{equation}
\label{musr_Lambda_LF_eqn}
\lambda_{LF}(\mu_0 H) = 2\Delta^2 \tau^x {\int_0}^\infty t^{-x} \exp(-\nu t)~ cos(2 \pi \mu_0 \gamma_\mu H t)~ dt
\end{equation}

this fit yields the fitting parameter as $x=0.76(2)$, $2 \Delta^2 \tau^x = 0.018(3)$~Hz and $ \nu_{T<10~K} = 2.04(7) \times 10^6$~Hz. This is lower than the high-temperature paramagnetic fluctuations, further corroborating the slowing of the spin fluctuations. Such a frequency range has been observed in other frustrated quantum spin liquid-like systems \cite{PhysRevLett.117.097201, PhysRevLett.125.267202, PhysRevB.110.L060403}.
For a conventional simple exponential auto-correlation function with a single-correlation time ($\tau$) yields a Lorentzian spectral density $S(\omega)$, which is related to the relaxation rate as given by, $1/\lambda_L \propto S(\omega)^{-1} \propto (\gamma_\mu \mu_0 H)^\eta$, with $\eta=2$ \cite{PhysRevB.20.850}. But at low temperature, SCNO exhibits $\eta=0.26(1)$, rather than $\eta=2$, indicating a more exotic distribution than the Lorentzian distribution. In general, $\eta \leq 1$ indicates a power-law behavior of the auto-correlation function, and such behaviors have been observed in different spin liquid systems \cite{PhysRevLett.96.247203, PhysRevLett.92.107204, PhysRevB.90.205103}. The asymmetry-time scaling behavior with the exponent $\eta = 0.26(1)$ as shown in the inset of Fig.\ref{fig1}(d) also corroborates the critical behavior expected for an RS phase. Therefore, on one hand, ZF measurements indicate the RS through the power-law behavior of the relaxation rate $\lambda_{ZF}$, on the other hand, LF measurements indicate the dynamic nature of the system. Hence, the $\mu$SR measurements conclusively suggest that the ground state of SCNO is a dynamical random singlet state.

\subsection{NMR} 
\begin{figure}
\includegraphics[scale=0.30]{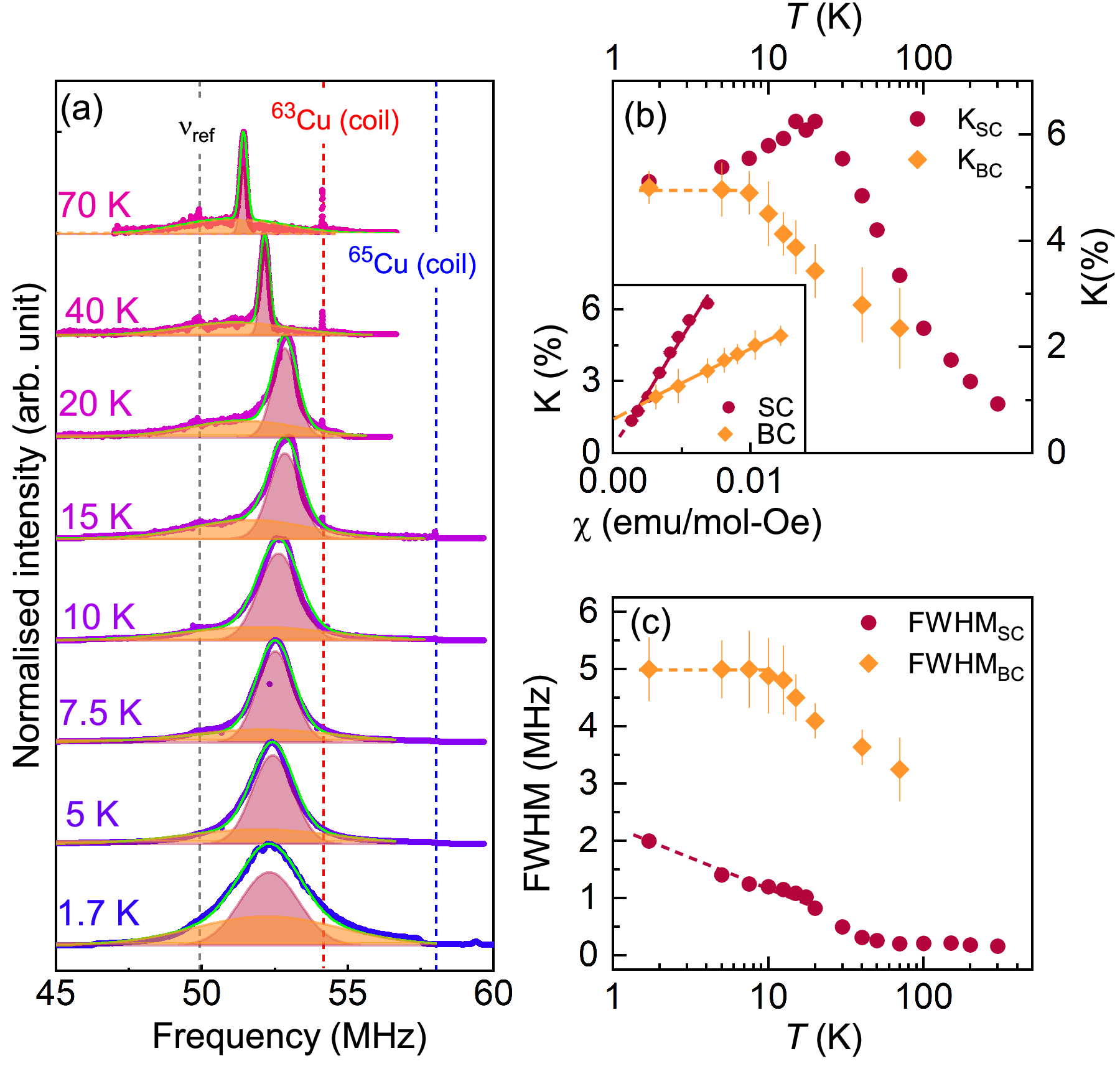}
\caption{\label{fig2}(a) Temperature evolution of $^{93}\mathrm{Nb}$ frequency sweep NMR spectra, measured at 4.8 T. The vertical black, red and blue dashed lines mark the Larmor frequencies for $^{93}\mathrm{Nb}$, $^{63}\mathrm{Cu}$ and $^{65}\mathrm{Cu}$ (contribution expected to be from the coil), respectively, at 4.8 T. The wine and orange highlighted areas represent the SC and BC components respectively in the total spectrum (green solid line). (b) Knight shift ($K$) as a function of temperature, for the two $^{93}\mathrm{Nb}$ components. Inset: $K$ versus susceptibility ($\chi$), considering temperature as an implicit parameter, for the two $^{93}\mathrm{Nb}$ components. The solid lines are a linear fit. (c) Temperature dependence of the full width at half maximum (FWHM) of the $^{93}\mathrm{Nb}$ NMR spectra, for the two $^{93}\mathrm{Nb}$ components.}
\end{figure}

After establishing a dynamical ground state with an inhomogeneous distribution of singlets in SCNO, we now turn to a further local probe, NMR. Owing to its inherently local character, NMR is sensitive to distinct local environments at each nuclear site and can thus distinguish their nature and their associated excitations, if any. Fig.~\ref{fig2}(a) illustrates the temperature dependence of the $^{93}\mathrm{Nb}$ NMR spectra acquired at a fixed field of 4.8 T. Even though $^{93}\mathrm{Nb}$ possesses a nuclear spin $I=9/2$, no satellite peaks are observed in the spectra. This absence is attributed to the highly symmetric local environment of the $^{93}\mathrm{Nb}$ ions, as revealed by the crystal structure \cite{PhysRevB.110.L020406}, which prevents the formation of a finite electric field gradient (EFG). Remarkably, the $^{93}\mathrm{Nb}$ NMR spectra reveal two distinct local magnetic environments associated with the central transition of $^{93}\mathrm{Nb}$, characterised by a sharp component (SC) and a broad background component (BC). At the reference frequency, indicated by the black dashed line in Fig.~\ref{fig2}(a), a temperature-independent peak is observed, which most likely arises from a minor impurity phase whose spectral weight is estimated to be below $2$\%. The red and blue dashed lines mark the $^{63}\mathrm{Cu}$ and $^{65}\mathrm{Cu}$ signals, respectively, originating from the copper NMR coil. The spectral profiles presented in Fig.~\ref{fig2}(a) were accurately modelled using two Gaussian lines, representing the distinct sharp and broad components inherent to the spectra. 

The temperature dependence of the Knight shifts for both components is illustrated in Fig.~\ref{fig2}(b). $K(T)$ is quantitatively described as $K(T) = \Delta \omega (T)/\omega_{L} \times 100\%$, where $\Delta \omega (T)$ represents the frequency deviation from the Larmor frequency, $\omega_{L}$. $K_{SC}$, exhibits a monotonic increase upon cooling down to approximately 20 K, below which it shows a gradual decrease, reflecting a non-monotonic temperature dependence of the local magnetic environment. This feature may indicate the formation of the RS state. One might ask whether a similar decrease of $K(T)$ could also be attributed to a gapped QSL. For a fully gapped QSL, however, the intrinsic spin susceptibility – and hence the NMR shift is expected to vanish as $T\rightarrow0$, because all spins are locked into singlets. Instead, in the RS state, a finite density of weakly bound or effectively free spins remains at low temperatures, so that after an initial decrease, $K(T)$ should approach a non-zero, temperature-independent value. In contrast to $K_{SC}$, $K_{BC}$ exhibits a steady increase with decreasing temperature until approximately $10$ K, where it begins to level off. The emergence of this temperature-independent plateau is characteristic of gapless magnetic excitations and aligns with observations in other gapless quantum spin liquid systems \cite{PhysRevB.96.174411, PhysRevB.111.L100403, PhysRevB.93.140408, PhysRevLett.119.227208}.

%%%%%%%%%%%%%%%%%%%%%%%%%%%%%%%%%%%%%%%%%%%%%%%%%%%%%%%%%%%%%%%%%%
%%%%%%%%%%%%%%%%%%%%%%%%%%%%%%%%%%%%%%%%%%%%%%%%%%%%%%%%%%%%%%%%%%

%%%%%%%%%%%%%%%%%%%%%%%%%%%%%%%%%%%%%%%%%%%%%%%%%%%%%%%%%%%%%%%%%%

\begin{figure*}
\includegraphics[scale=0.35]{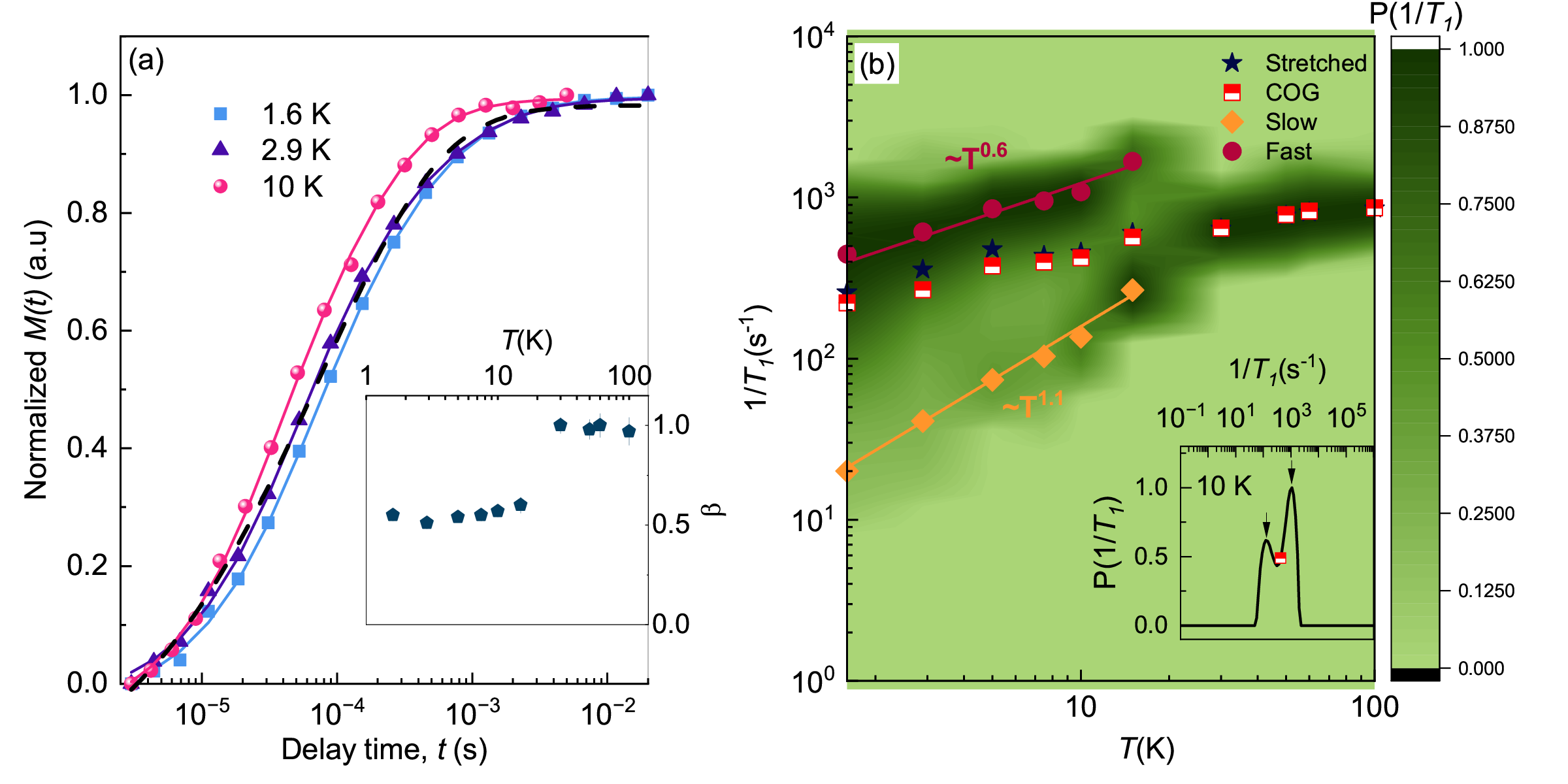}
\caption{\label{fig3}(a) $^{93}\mathrm{Nb}$ NMR normalized magnetization recovery curves, $M(t)$, at selected temperatures, fitted with stretched exponential function (black dashed line) and ILT analyses (solid lines). Inset: Temperature dependence of the stretched exponent ($\beta_{n}$), extracted from fits to the $^{93}\mathrm{Nb}$ NMR relaxation data. (b) A color contour plot of $1/T_{1}$ as a function of temperature $T$, where the intensity represents the probability distribution $P(1/T_{1})$. The temperature dependence of the spin–lattice relaxation rate ($1/T_{1}$) for both the slow and fast components, extracted from the peak positions, follows power-law behaviour (solid lines). Inset: $P(1/T_{1})$ obtained from ILT analysis at $10$ K with arrows indicating the $1/T_{1}$ values of the slow and fast components.}
\end{figure*}

%%%%%%%%%%%%%%%%%%%%%%%%%%%%%%%%%%%%%%%%%%%%%%%%%%%%%%%%%%%%%%%%%%

From Fig.~\ref{fig2}(c), it is evident that $\mathrm{FWHM}_{BC}$ is approximately three times greater than that of $\mathrm{FWHM}_{SC}$. This behaviour is naturally explained by a substantially shorter spin-spin relaxation time ($T_{2}$) for BC, which progressively broadens its lineshape upon cooling relative to SC. $\mathrm{FWHM}_{SC}$ exhibits a pronounced increase upon cooling, indicative of diverging spectral broadening at low temperature, while $\mathrm{FWHM}_{BC}$ tends to saturate below $10$ K, corroborating the temperature dependence of $K_{BC}$. 

The hyperfine coupling constants for SC and BC denoted by $A_{hf}^{SC}$ and $A_{hf}^{BC}$ respectively, depicted in the inset of Fig.~\ref{fig2}(b), are estimated using the relation 
\begin{equation}
\label{NMR_eqn1}
K = K_{0} + (A_{hf}/N_{A}\mu_{B})\chi(T),
\end{equation}
where $K_{0}$ represents the temperature-independent contribution to the Knight shift, $N_{A}$ is Avogadro's number and $\chi(T)$ is the bulk magnetic susceptibility. The resulting values, $A_{hf}^{SC}=5.17~ T/\mu_{B}$ and $A_{hf}^{BC}=1.59 ~ T/\mu_{B}$, imply an approximately $3.2$-times enhancement of the hyperfine coupling for SC relative to BC, a factor that provides a key constraint on the microscopic assignment of the two components as discussed in the following sections. We note that, in a structurally disordered system such as SCNO, where Cu/Nb site mixing is present, it is natural to expect multiple local Nb environments, which may modify the local hyperfine coupling $A_{hf}$ and, consequently, the NMR line shape and $1/T_{1}$. In this sense, the two components SC and BC can be viewed as representative Nb probes of two classes of local environments: one more strongly coupled to Cu spins, with enhanced $A_{hf}$, stronger low temperature broadening; and another more weakly coupled, with reduced $A_{hf}$, saturating $K$ and FWHM.   

To further characterize the dynamics of the two spectral components, spin-lattice relaxation rates ($1/T_{1}$) were measured using saturation recovery at multiple temperatures, with data acquired specifically at the peak position of SC. Since these measurements target the common spectral maximum of the overlapping line-shapes, they predominantly sample nuclei contributing to the dominant detected intensity across the relevant temperature range. We thus consider the extracted relaxation-rate distribution ($P(1/T_{1})$) as representative of the central resonance region, despite not constituting a complete frequency-averaged distribution across the full linewidth. In general, $1/T_{1}$ is related to the imaginary part of dynamical spin susceptibility, $\chi^{\prime \prime}(q,\omega)$, by the relation,
\begin{equation}
\label{NMR_eqn2}
\frac{1}{T_{1}T} \sim  \sum_{q} \left| A_{hf}(q)\right|^2 \frac{\chi^{\prime \prime}(q,\omega\rightarrow 0)}{\omega}.
\end{equation} 
In Fig.~\ref{fig3}(a), the magnetization recovery curves, $M(t)$, at selected temperatures are summarized. Since a single exponential fit was inadequate owing to variations in the magnetic environment producing a distribution of $1/T_{1}$ relaxation rates, the recovery curves were fitted using a stretched exponential function for $I=9/2$ [black dashed line in Fig.~\ref{fig3}(a)] described by the equation \cite{PhysRevB.96.174411},
\begin{equation}
\begin{aligned}
1 - M(t)/M(\infty) = C \Big(&
0.006 e^{-(t/T_{1,str})^{\beta}} + 0.0335 e^{-(6t/T_{1,str})^{\beta}} \\
&+ 0.0925 e^{-(15t/T_{1,str})^{\beta}} \\
&+ 0.215 e^{-(28t/T_{1,str})^{\beta}} \\
&+ 0.653 e^{-(45t/T_{1,str})^{\beta}}
\Big),
\end{aligned}
\end{equation}
where $C$ is a pre-factor and $\beta$, known as the stretched exponent, assumes a value less than $1$, reflecting deviation from single-exponential relaxation. The temperature dependence of $\beta$, shown in the inset of  Fig.~\ref{fig3}(a), demonstrates that a stretched exponential fit is particularly necessary below $30$ K.

%\textcolor{red}{It is noteworthy that NMR in disordered systems probes an ensemble-averaged nuclear response, biased toward slow, long-time spin fluctuations with a broad distribution of relaxation rates, yielding $\beta_{n}< 1$. In contrast, $\mu$SR depolarization is governed by stochastic sporadic bursts of strong local fields from intermittent spin dynamics within the muon finite lifetime, resulting in rapid early-time decay, hence $\beta_\mu > 1$} \cite{PhysRevLett.127.157204, slichter1996}.

Furthermore, the recovery curves were analyzed using an inverse Laplace transform (ILT) approach with Tikhonov regularization (solid lines in Fig.~\ref{fig3}(a)) \cite{SONG2002261, PhysRevB.101.174508, Wang2021}, which explicitly accounts for the presence of multiple relaxation channels. This procedure yields a significantly improved description of the experimental data compared to the conventional stretched-exponential fit, as illustrated in Fig.~\ref{fig3}(a). The ILT analysis fits the measured inversion recovery curve $M(t_{i})$ (at times $t_{i}$) to a multi-exponential summed over discrete $1/T_{1j}$ with weights $P(1/T_{1j})$:
\begin{equation}
\label{NMR_eqn4}
M(t_{i})=\sum_{j=1}^{m} \sum_{k=1}^{5} [1-2p_{k}e^{-q_{k}t_{i}/T_{1j}}]P(1/T_{1j})+E(t_{i}),
\end{equation}
where, for $I=9/2$, the coefficients are $p_{k}=\lbrace 0.006, 0.0335, 0.0925, 0.215, 0.653\rbrace$ and $q_{k}=\lbrace 1, 6, 15, 28, 45\rbrace$ with $\sum_{k=1}^{5} p_{k}=1$ and $E(t_{i})$ is the experimental noise. The ILT analysis yields $P(1/T_{1})$ at each temperature, with an exemplary result at $T = 10$ K shown in the inset of Fig.~\ref{fig3}(b). This distribution is convoluted into two distinct channels- fast and slow, with the corresponding $1/T_{1}$ values determined from the peak positions (arrows). Fig.~\ref{fig3}(b) presents a colour contour plot of $1/T_{1}$ versus temperature $T$, where the colour intensity reflects $P(1/T_{1})$, along with the temperature dependence of both relaxation channels extracted from the ILT peaks and their center-of-gravity (COG) values. In particular, the COG aligns closely with the $1/T_{1,str}$ values derived from stretched exponential fitting. Notably, the presence of these two separate relaxation channels could not be discerned from the stretched exponential analysis, thus underscoring the essential utility of ILT in revealing this dynamic heterogeneity. 

Both $1/T_{1}^{fast}$ and $1/T_{1}^{slow}$ follow power-law temperature dependences, scaling as $T^{0.6}$ and $T^{1.1}$ respectively. Notably, their ratio, $(1/T_{1}^{fast})/(1/T_{1}^{slow})\approx 10$ (Fig.~\ref{fig3}(b)) aligns quantitatively with the ratio of hyperfine coupling constants extracted from the spectra, $A_{hf}^{SC}/A_{hf}^{BC} \approx 3.2$ (whose square is of the order of $10$), as anticipated from Eq.(\ref{NMR_eqn2}). This establishes that the fast and slow relaxation channels correspond directly to the SC and BC spectral components, respectively. We now focus on the nature of the individual components:
\\(i) As noted earlier, both $K_{BC}$ and $\mathrm{FWHM}_{BC}$ saturate at low temperatures (Fig.~\ref{fig2}(b,c)), while $1/T_{1}^{slow}$ (Fig.~\ref{fig3}(b)) obeys a power law consistent with a gapless state potentially hosting a quantum spin liquid (QSL) like ground state, as observed in other frustrated quantum magnets \cite{PhysRevB.96.174411, PhysRevB.111.L100403, PhysRevB.93.140408, PhysRevLett.119.227208}. 
\\(ii) In contrast, $K_{SC}$ decreases below $\sim 20$ K, signalling the development of a spin-singlet-like ground state, while $\mathrm{FWHM}_{SC}$ exhibits a pronounced low-temperature upturn, characteristic of RS behaviour \cite{PhysRevB.107.014424}. Consistently, $1/T_{1}^{\mathrm{fast}}$ also follows a power-law temperature dependence, as expected for an RS system, which is gapless and therefore exhibits a power-law behaviour in $1/T_{1}$ \cite{rw5c-491w,chen2025spinonssolitonsrandomsinglets}. 

Several disorder-driven systems proposed to host RS physics exhibit dynamical phase separation, manifested as two distinct relaxation channels in $1/T_{1}$ \cite{Wang2021, PhysRevB.110.064418, PhysRevB.107.014424, chen2025spinonssolitonsrandomsinglets}. However, there are also RS systems in which such dynamical phase separation has not been reported \cite{rw5c-491w, PhysRevB.102.094407}. One possible reason is that ILT analysis was not performed in those studies, which may limit the ability to resolve multiple relaxation channels. In addition, the choice of experimental parameters, particularly the pulse separation ($\tau$) can play a crucial role for the observation of static phase separation, seen in spectral analysis. If $\tau$ is not sufficiently short, components with very fast $T_{2}$ may decay too rapidly to be detected, thereby masking the presence of additional phases in the spectra \cite{Wang2021, PhysRevB.110.064418, PhysRevB.107.014424, chen2025spinonssolitonsrandomsinglets, PhysRevB.102.094407}. Therefore, the absence of reported phase separation in some RS systems does not necessarily rule out its existence. Rather, it may reflect limitations in experimental sensitivity. In this context, our observation of two distinct components, supported by both spectral analysis and ILT results, provides strong evidence for phase separation. While such coexistence appears to be a natural consequence of disorder in RS systems, its experimental visibility depends sensitively on the measurement protocol.

It is noteworthy that a recent study on a sister compound reported three characteristic energy scales, where the exchange randomness is broad but bounded, and the probability distribution, $P(J)$ deviates from the conventional power-law form expected for an RS state \cite{mahapatra2026emergentrandomspinsinglets}. In that case, $P(J)$ follows a power-law behavior only over a limited temperature range. A similar scenario may apply to our system, as both $\mu$SR and NMR indicate a characteristic scale near $20$ K, although additional energy scales may lie beyond our experimental window. Consistently, another analogous compound near the site-percolation threshold has recently been reported to exhibit coexistence of distinct spin environments based on experimental and theoretical studies, supporting our results \cite{xu2026dynamicalmagnetismdisorderedcubic}. Recent theoretical studies of the disordered $J_{1}-J_{2}$ Heisenberg model on the square lattice suggest the possible existence of the valence bond glass (VBG) ground state \cite{dash2026valencebondglassglassy}. However, this scenario can likely be ruled out for our system, since VBG states are typically associated with a pseudo-gap \cite{Tarzia_2008, PhysRevB.103.224430, Mustonen2022} that is not observed in our data, atleast within our experimental resolution. 

%%%%%%%%%%%%%%%%%%%%%%%%%%%%%%%%%%%%%%%%%%%%%%%%%%%%%%%%%%%%%%%%%%%%%%%%%%%%%%%%%%%%%%%%%%%%%%%%%%%%%%%%%%%%%%%%%%%%%%%%%%%%%%%%%%%%%%%%%%%%%%%%%%%%%%%%%%%%%%%%%%%%%%%%%%%%%%%%%%%%%%%%%%%%%%%%%%%%%%%%%%%%%%%%%%%%%%%%%%%%%%%%%%%%%%%%%
\subsection{CONCLUSION} Utilizing two complementary microscopic probes, $\mu$SR and NMR, the presence of two distinct dynamical magnetic states has been observed in SCNO. The $\mu$SR spectroscopy analysis indicates the RS-singlet ground state owing to the power-law of the relaxation rates as a function of temperature and fields. The dynamical behavior was also confirmed by the lack of muon-spin decoupling under various longitudinal fields. 

The $^{93}$Nb NMR spectra clearly resolve two components, SC and BC, which exhibit markedly distinct temperature dependence in both $K(T)$ and FWHM, indicating microscopically distinct local magnetic environments. Conventional stretched-exponential fits to the spin-lattice relaxation recovery curves fail to disentangle the contributions from these components, necessitating an ILT analysis. The ILT reveals two well-separated peaks in the relaxation-rate distribution $P(1/T_{1})$ corresponding to fast and slow channels that follow distinct power-law temperature dependence ($T^{0.6}$ and $T^{1.1}$, respectively). A combined analysis of the spectral features and relaxation data supports the assignment of the fast relaxation channel to SC (RS-like) and the slow relaxation channel to BC (QSL-like). More generally, in disorder-driven systems exhibiting RS physics, both static and dynamical inhomogeneities are expected, which can manifest as phase separation in local probes such as NMR.

%%%%%%%%%%%%%%%%%%%%%%%%%%%%%%%%%%%%%%%%%%%%%%%%%%%%%%%%%%%%%%%%%%%%%%%%%%%%%%%%%%%%%%%%%%%%%%%%%%%%%%%%%%%%%%%%%%%%%%%%%%%%%%%%%%%%%%%%%%%%%%%%%%%%%%%%%%%%%%%%%%%%%%%%%%%%%%%%%%%%%%%%%%%%%%%%%%%%%%%%%%%%%%%%%%%%%%%%%%%%%%%%%%%%%%%%%

\subsection{ACKNOWLEDGEMENT}
K.M.R. acknowledges the financial support by the Deutsche Forschungsgemeinschaft  (DFG, German Research Foundation)  under Germany’s Excellence Strategy through the Würzburg-Dresden Cluster of Excellence on Complexity and Topology in Quantum Matter-ct.qmat (EXC 2147, project-id 390858490). M.N. acknowledges financial support from the Deutsche Forschungsgemeinschaft via the Walter Benjamin Program (Grant NA 2012/1-1, Project No.~540160192).

\bibliography{references}
\end{document}